\documentclass[12pt]{article}
\usepackage{amssymb,amsfonts}
\def \la{\lambda}

\def \sh{{\rm sinh}}

\def \t{\tau}

\newcounter{map}
\newcounter{fel}
\newcounter{aw}
\newcounter{aww}
\begin{document}
\baselineskip 18pt

\title{Defining relations on the Hamiltonians of $XXX$ and $XXZ$ $R$-matrices and
new integrable spin-orbital chains}
\author{P.~N.~Bibikov\\V.~A.~Fock Institute of Physics,
Sankt-Petersburg State University}

\maketitle

\vskip5mm

\begin{abstract}
Several complete systems of integrability conditions on a spin chain Hamiltonian density
matrix are presented. The corresponding formulas for $R$-matrices are also given.
The latter is expressed via the local Hamiltonian density in the form similar to spin one half
$XXX$ and $XXZ$ models. The result is applied to the problem of integrability of
$SU(2)\times SU(2)$- and $SU(2)\times U(1)$-invariant spin-orbital chains
(the Kugel-Homskii-Inagaki model). The eight new integrable cases are found. One of them
corresponds to the Temperley-Lieb algebra,
the others three to the algebra associated with the $XXX$, $XXZ$ and graded $XXZ$
models. The last two $R$-matrices are also presented.
\end{abstract}

\section{Hamiltonian density and $R$-matrix}

It is well known \cite{1},\cite{3}, that an $R$-matrix, which satisfy the Yang-Baxter equation
(in the braid group form),
\begin{equation}
R_{12}(\la-\mu)R_{23}(\la)R_{12}(\mu)=R_{23}(\mu)R_{12}(\la)R_{23}(\la-\mu),
\end{equation}
(usually denoted by $\check R(\lambda)$) and the initial condition,
\begin{equation}
R(0)\propto I,
\end{equation}
where $I$ is an identity matrix, produces according to the formula:
\begin{equation}
H=\frac{\partial}{\partial\lambda}R(\lambda)|_{\lambda=0},
\end{equation}
the corresponding to an integrable spin chain Hamiltonian density. Let us remain that both $H$,
and $R(\lambda)$ are $M^2\times M^2$ ($M=2,3,4,...$) matrices. For such a matrix
$X$ the corresponding $M^3\times M^3$ matrices $X_{12}$ and $X_{23}$ are defined as follows:
\begin{equation}
X_{12}=X\otimes I_M,\quad X_{23}=I_M\otimes X.
\end{equation}
Here $I_M$ is the identity  $M\times M$ matrix. The Hamiltonian may be obtained according to the
formula:
\begin{equation}
\hat H=\sum_nH_{n,n+1}.
\end{equation}

A matrix $R(\lambda)$ is important for applications in statistical physics. However for the
problems related to spin chains just the matrix $H$ is a starting point. By this reason it
is necessary to be able to reverse the formula (3) and get the $R$-matrix from a Hamiltonian
density.
The system of necessary conditions for the such operation is an infinite sequence of integrability
conditions. The first one (the Reshetikhin condition)
\begin{equation}
{[}H_{12}+H_{23},{[}H_{12},H_{23}{]]}=K_{23}-K_{12},
\end{equation}
is known long ago \cite{4}, the next two are presented in \cite{5},\cite{6}, while the others are
not yet obtained. Let us notice that the condition (6) will be satisfied when it will be found an
appropriate matrix $K$ for the right side.

In \cite{6} the system of integrability conditions was studied in the context of a power expansion
of a $R$-matrix (1)-(3) with respect to $\lambda$. A new method for solving
the Yang-Baxter equation based on an analysis of such expansion was also presented in this paper.
This method turned out to be rather effective. In particular it allowed to find all
$R$-matrices for integrable spin-ladders \cite{7}. At the same time taking into account only the
{\it necessary} integrability conditions this method does not guarantee an existance of a solution.
Consiquent solution of the integrability conditions together with expanding of quotents for the
$R$-matrix entries into power series gives a possibility at a some stage to suggest their
explicit forms. The latter may be expressed in the framework of rational, trigonometric
(exponential) or elliptic functions. Nevertheless the Eq. (1) must be checked for the obtained
$R$-matrix.

From the other hand it is better to the reverse the formula (3) having in a stock of knowledge a
rather big number of {\it sufficient} integrability conditions, related to explicit formulas
expressing an $R$-matrix from the corresponding Hamiltonian density. As an example we may suggest
the sufficient condition virtually obtained in \cite{8} and related to the Temperley-Lieb algebras.
\cite{9}. Namely if the matrix $H$ satisfy the following system conditions:
\begin{equation}
H^2=-dH,\quad
H_{12}H_{23}H_{12}=H_{12},\quad
H_{23}H_{12}H_{23}=H_{23},
\end{equation}
then the Eq. (6) turns into identity for $K=2H$. Moerover the following
$R$-matrix,
\begin{equation}
R^{TL}(\lambda)=I{\rm sh}(\lambda+\eta)+H{\rm sh}\lambda,
\end{equation}
where ${\rm ch}\eta=d/2$ according to (3) generates a Hamiltonian density, that differs from $H$
by the item $I{\rm ch}\eta$, which is proportional to the identity matrix and results to an
unsufficient shift on a constant in the spectrum of the Hamiltonian.

In the next sections we shall represent the two more sets of sufficient conditions on the
Hamiltonian density. These conditions were obtained by studying the structures of
$R$-matrices for $XXX$, $XXZ$ and graded $XXZ$ models of half spin.

\section{The algebras related to Hamiltonian densities}

The following system of relations (satisfied for the $XXX$ model),
\begin{equation}
H^2=\propto I,\quad
H_{12}H_{23}H_{12}=H_{23}H_{12}H_{23},
\end{equation}
is related to the $R$-matrix:
\begin{equation}
R^{XXX}(\lambda)=\eta I+\lambda H.
\end{equation}

The $R$-matrices for $XXZ$ and graded $XXZ$ models are expressed from the related Hamiltonian
densities according to the formula:
\begin{equation}
R^{XXZ}(\lambda)=H{\rm sh}\lambda+\frac{{\rm ch}\lambda}{{\rm sh}\eta}(H^2-I)+
\frac{1}{{\rm sh}\eta}({\rm ch}^2\eta I-H^2).
\end{equation}
In this case the matrix $H$ depends on the real parameter $a={\rm ch}\eta$ (the case $a<1$
corresponds to the $\sin-\cos$ $R$-matrix \cite{3}), and satisfy a system of relations which
generalises of (9). First of all it includes the Yang-Baxter equation for the Hamiltonian
density,
\begin{equation}
H_{12}H_{23}H_{12}=H_{23}H_{12}H_{23},
\end{equation}
the relation
\begin{equation}
(H_{12}-H_{23})^3=(a^2+2)(H_{12}-H_{23}),
\end{equation}
and the following set of relations,
\begin{eqnarray}
H_{12}H_{23}(H_{12}^2-I)&=&(H_{23}^2-I)H_{12}H_{23},\nonumber\\
H_{12}(H_{23}^2-I)H_{12}&=&H_{23}(H_{12}^2-I)H_{23},\nonumber\\
(H_{12}^2-I)H_{23}H_{12}&=&H_{23}H_{12}(H_{23}^2-I),\nonumber\\
H_{12}(H_{23}^2-I)(H_{12}^2-I)&=&(H_{23}^2-I)(H_{12}^2-I)H_{23},\nonumber\\
(H_{12}^2-I)H_{23}(H_{12}^2-I)&=&(H_{23}^2-I)H_{12}(H_{23}^2-I),\nonumber\\
(H_{12}^2-I)(H_{23}^2-I)H_{12}&=&H_{23}(H_{12}^2-I)(H_{23}^2-I),\nonumber\\
(H_{12}^2-I)(H_{23}^2-I)(H_{12}^2-I)&=&(H_{23}^2-1)(H_{12}^2-I)(H_{23}^2-I),
\end{eqnarray}
appearing from a direct substitution of (11) into (1).

Evidently the system (14) is overfulled and we have written it only to make clear the proof of
the Eq. (1) for the $R$-matrix (11). However it is not so obvious but in this case the
Eq. (13) is necessary as well as (12) and (14).
The following two conditions are also satisfied for the standard $S=1/2$ $XXZ$ $R$-matrix:
\begin{eqnarray}
H_{12}^2H_{23}^2&=&H_{23}^2H_{12}^2,\\
H^3&=&aH^2+H-aI.,
\end{eqnarray}
Moreover the system (14) may be easily proved from (12), (13), (15), (16).
For the $R$-matrix of the graded $XXZ$ model \cite{3} the relations (12)-(15) are still satisfied,
but the Eq. (16) substitutes by a more general one:
\begin{equation}
H^4=(a^2+1)H^2-a^2I.
\end{equation}

In this case an appropriate choice of the minimal set of relations on a Hamiltonian density still
remains an open question.

\section{Hamiltonian densities and related $R$-matrices for spin-orbital chains}

The following Hamiltonian density was suggested in \cite{10}-\cite{12} in order to describe
magnetic systems with an additional orbital degeneracy in electron systems of atoms,
\begin{eqnarray}
H_{n,n+1}&=&(s_ns_{n+1})(a_1+a_2(\t^x_n\t^x_{n+1}+\t^y_n\t^y_{n+1})+a_3\t^z_n\t^z_{n+1}+
\frac{1}{2}a_6(\tau_n^z+\tau_{n+1}^z))+\nonumber\\
&+&a_4(\t_n^x\t_{n+1}^x+\t_n^y\t_{n+1}^y)
+a_5\t^z_n\t^z_{n+1}+\frac{1}{2}a_7(\tau_n^z+\tau_{n+1}^z).
\end{eqnarray}
Here the spin and pseudospin operators $s_n$ and $\tau_n$ are related to the Pauli matrices,
\begin{equation}
s^k=\frac{1}{2}\sigma^k\otimes I_2,\quad \tau^k=\frac{1}{2}I_2\otimes\sigma^k,\quad k=x,y,z.
\end{equation}
In below the set of coefficients $\{a_1,..a_7\}$ will be denoted by $S$.
The martrix
$H\in {\rm End}(({\mathbb C}^2\otimes{\mathbb C}^2)\otimes({\mathbb C}^2\otimes{\mathbb C}^2))$
related to (18) is invariant under the following symmetry,
\begin{equation}
H\rightarrow Q\otimes QHQ^{-1}\otimes Q^{-1},
\end{equation}
where $Q\in SU(2)\times U(1)$ and the group $U(1)$ is generated by the matrix $\tau^z$.

The following substitution of coefficients:
\begin{equation}
\{a_1,a_2,a_3,a_4,a_5,a_6,a_7\}\rightarrow\{a_1,a_2,a_3,a_4,a_5,-a_6,-a_7\},
\end{equation}
relates to the transformation (20) with $Q=I_2\otimes\sigma^x$ and
therefore does not change the spectrum of the Hamiltonian as well as does not fail the
integrability.

The similar is true for the following substitution:
\begin{equation}
\{a_1,a_2,a_3,a_4,a_5,a_6,a_7\}\rightarrow\{a_1,-a_2,a_3,-a_4,a_5,a_6,a_7\},
\end{equation}
related to the symmetry (20)
\begin{equation}
H_{2n,2n+1}\rightarrow4\tau_{2n}^zH_{2n,2n+1}\tau_{2n}^z,\quad
H_{2n-1,2n}\rightarrow4\tau_{2n}^zH_{2n-1,2n}\tau_{2n}^z.
\end{equation}
for a chain with even number of sites.

The only one integrable case of the model (18) was known up to now. It relates to
$S_0=\{1,4,4,1,1,0,0\}$ \cite{13}-\cite{15}. In this, $SU(4)$-case the shifted on
$1/4I$ Hamiltonian density is equal to the permutation operator in
${\mathbb C}^4\otimes{\mathbb C}^4$, so all the conditions (9) are satisfied, and the $R$-matrix
relates to the forlmula (10). Let us notice that the dual model obtained by the transformation (22)
is also integrable with the $R$-matrix related to the same formula.
A modern discussion of the $SU(4)$-symmetric model is given in \cite{16},\cite{17}.
The $SU(2)\times SU(2)$- and $SU(2)\times U(1)$-symmetric
models were also discussed in \cite{18},\cite{19}.

After substitution of (18) into the condition (6) we have obtained 8 new solutions.
($\varepsilon,\varepsilon_1,\varepsilon_2=\pm1$),
\begin{eqnarray}
S_1&=&\{1,-4\varepsilon,-4,\varepsilon,1,0,0\},\\
S_2&=&\{0,4,0,1,0,4\varepsilon,\varepsilon\},\\
S_3&=&\{1,8\varepsilon_1,4,2\varepsilon_1,3,4\varepsilon_2,-\varepsilon_2\},\\
S_4&=&\{1,8\varepsilon_1,4,2\varepsilon_1,-1,4\varepsilon_2,3\varepsilon_2\},\\
S_5&=&\{0,4,0,1,\alpha,0,0\},\\
S_6&=&\{0,4,0,1,0,0,\alpha\},\\
S_7&=&\{1,4\varepsilon,8,2\varepsilon,1,0,0\},\\
S_8&=&\{2,4\varepsilon,-8,-\varepsilon,5,0,0\}.
\end{eqnarray}

In the case $S_1$ the matrix $H-1/4I$ satisfy the Temperley-Lieb condition (7) ($d=4$).
and has the following form \cite{8}:
\begin{equation}
H-\frac{1}{4}I=b\otimes\bar b,
\end{equation}
(or in component notation $H_{ij,kl}=b_{ij}\bar b_{kl}$). Bivectors $b$ and $\bar b$,
written in (32) as a column and row, is more convenient to represent as matrices,
\begin{equation}
b=\left(\begin{array}{cccc}
0&0&0&1\\
0&0&-\varepsilon&0\\
0&-1&0&0\\
\varepsilon&0&0&0
\end{array}\right),\quad
\bar b=\left(\begin{array}{cccc}
0&0&0&-1\\
0&0&\varepsilon&0\\
0&1&0&0\\
-\varepsilon&0&0&0
\end{array}\right),
\end{equation}
satisfying the following condition,
\begin{equation}
b\bar b=-\varepsilon I.
\end{equation}

The $R$-matrix related to $S_1$
has the form (8). For $\varepsilon=1$ this model is $SU(2)\times SU(2)$-symmetric.
As it was mentioned in \cite{18}, the corresponding ferromagnetic model
(with the Hamiltonian density $-H$) corresponds to the critical point, separating 4 different
phases. The ground state of this model is highly degenerate.

In the case $S_2$ the matrix $H$ satisfies the condition (9), and therefore the
$R$-matrix is given by the formula (10). In the case $S_3$ the shifted matrix
$\tilde H=H+\frac{3}{4}I$ satisfy (9) and the related $R$-matrix is also repersented by (10).
In the case $S_4$ the shifted matrix $\tilde H=H+\frac{1}{4}I$ satisfies (9) and its $R$-matrix
also relates to (10).

In the case $S_5$
the shifted matrix $\tilde H=H+\frac{\alpha}{4}I$ satisfies (12)-(16) with $a=\frac{\alpha}{2}$
and therefore the related $R$-matrix is given by (11). In the case $S_6$ the Hamiltonian density
satisfies (12)-(15), (17) and the related $R$-matrix also has the form (11).

Finitely we represent the $R$-matrices, related to $S_7$ and $S_8$,
\begin{equation}
R^{(7)}=\left(\begin{array}{cccccccccccccccc}
f_1&0&0&0&0&0&0&0&0&0&0&0&0&0&0&0\\
0&f_2&0&0&\varepsilon g_1&0&0&0&0&0&0&0&0&0&0&0\\
0&0&f_2&0&0&0&0&0&g_1&0&0&0&0&0&0&0\\
0&0&0&f_3&0&0&\varepsilon\lambda&0&0&-\lambda&0&0&\varepsilon g_2&0&0&0\\
0&\varepsilon g_1&0&0&f_2&0&0&0&0&0&0&0&0&0&0&0\\
0&0&0&0&0&f_1&0&0&0&0&0&0&0&0&0&0\\
0&0&0&\varepsilon\lambda&0&0&f_3&0&0&\varepsilon g_2&0&0&-\lambda&0&0&0\\
0&0&0&0&0&0&0&f_2&0&0&0&0&0&g_1&0&0\\
0&0&g_1&0&0&0&0&0&f_2&0&0&0&0&0&0&0\\
0&0&0&-\lambda&0&0&\varepsilon g_2&0&0&f_3&0&0&\varepsilon\lambda&0&0&0\\
0&0&0&0&0&0&0&0&0&0&f_1&0&0&0&0&0\\
0&0&0&0&0&0&0&0&0&0&0&f_2&0&0&\varepsilon g_1&0\\
0&0&0&\varepsilon g_2&0&0&-\lambda&0&0&\varepsilon\lambda&0&0&f_3&0&0&0\\
0&0&0&0&0&0&0&g_1&0&0&0&0&0&f_2&0&0\\
0&0&0&0&0&0&0&0&0&0&0&\varepsilon g_1&0&0&f_2&0\\
0&0&0&0&0&0&0&0&0&0&0&0&0&0&0&f_1
\end{array}\right),
\end{equation}
£¤¥ $f_1=(1+\la)(1+3\la),\quad f_2=1+\la,\quad f_3=1+2\la,\quad
g_1=3\la(1+\la),\quad g_2=\la(2+3\la)$ ¨
\begin{equation}
R^{(8)}=\left(\begin{array}{cccccccccccccccc}
f_1&0&0&0&0&0&0&0&0&0&0&0&0&0&0&0\\
0&f_2&0&0&0&0&0&0&0&0&0&0&0&0&0&0\\
0&0&f_1&0&0&0&0&0&0&0&0&0&0&0&0&0\\
0&0&0&f_3&0&0&-g_1&0&0&g_2&0&0&g_1&0&0&0\\
0&0&0&0&f_2&0&0&0&0&0&0&0&0&0&0&0\\
0&0&0&0&0&f_1&0&0&0&0&0&0&0&0&0&0\\
0&0&0&-g_1&0&0&f_3&0&0&g_1&0&0&g_2&0&0&0\\
0&0&0&0&0&0&0&f_1&0&0&0&0&0&0&0&0\\
0&0&0&0&0&0&0&0&f_1&0&0&0&0&0&0&0\\
0&0&0&g_2&0&0&g_1&0&0&f_3&0&0&-g_1&0&0&0\\
0&0&0&0&0&0&0&0&0&0&f_1&0&0&0&0&0\\
0&0&0&0&0&0&0&0&0&0&0&f_2&0&0&0&0\\
0&0&0&g_1&0&0&g_2&0&0&-g_1&0&0&f_3&0&0&0\\
0&0&0&0&0&0&0&0&0&0&0&0&0&f_1&0&0\\
0&0&0&0&0&0&0&0&0&0&0&0&0&0&f_2&0\\
0&0&0&0&0&0&0&0&0&0&0&0&0&0&0&f_1
\end{array}\right).
\end{equation}
Here $g_2=f_2-f_3$ and $f_1=f_2{\rm e}^{\lambda}=4e^{2\la}-1,\quad f_3=2e^{\la}+e^{-\la},
\quad g_1=\varepsilon(1-e^{-2\la}),\quad g_2=4\sh\la$.
A dependence of these matrices on $H$ is not yet found.

\section{Conclusions}

The two sets of sufficient conditions of integrability and related explicit formulas
experssing the $R$-matrices from the corresponding Hamiltonian densities are presented.
These conditions were obtained from an analysis of the $R$-matrices and related Hamiltonian
densities for the $XXX$ and $XXZ$ models of half spin. The suggested approach develops an idea
virtually used by P.~P.~Kulish for the models related to Temperley-Lieb algebras.
The new method was applied for obtaining of the new $SU(2)\times SU(2)$-invariant and 7 new
$SU(2)\times U(1)$-invariant $R$-matrices for spin-orbital chains.

The author is very grateful to P.~P.~Kulish for the stimulating discussion and helpful comments.

\end{document}